\documentclass{naturef}
\usepackage{graphicx}
\usepackage{epsfig}
\def\ga{\mathrel{\hbox{\rlap{\hbox{\lower4.5pt\hbox{$\sim$}}}{\raise1pt\hbox{$>$}}}}}
\def\la{\mathrel{\hbox{\rlap{\hbox{\lower4.5pt\hbox{$\sim$}}}{\raise1pt\hbox{$<$}}}}}

\title{Hard X-ray emission lines from the decay of {\boldmath
    $^{44}$}Ti  in the remnant of
  supernova~1987A\footnote{published in {\it Nature\/}, 490,
    373-375 (2012).}}  

\author{S. A. Grebenev$^{1}$, A. A. Lutovinov$^{1}$, S. S. Tsygankov$^{1,2,3,4}$
\& C. Winkler$^{5}$}

\begin{document}

\maketitle
\baselineskip 20pt
\parsep 6pt
\parskip 6pt

\begin{affiliations}
 \item Space Research Institute, RAS,
   Profsoyuznaya 84/32, 117997 Moscow,
   Russia.
 \item Max-Planck-Institut\,f\"ur\,Astrophysik,\,Karl-Schwarzschild-Strasse\,1,\,D-85741\,Garching,\,Germany. 
 \item FINCA, University of Turku, V\"ats\"al\"antie 20,
   FI-21500 Piikki\"o, Finland.
 \item Astronomy Division, Department of Physics, University of
   Oulu, FI-90014 Oulu, Finland.
 \item European Space Agency, ESTEC, Keplerlaan
   1, 2200 AG Noordwijk, The Netherlands.
\end{affiliations}

\begin{abstract}
It is assumed\cite{suntzeff92,clayton92,fransson93} that the
radioactive decay of $^{\bf 44}$Ti powers the infrared, optical
and UV emission of supernova remnants after the complete decay
of $^{\bf 56}$Co and $^{\bf 57}$Co (the isotopes that dominated
the energy balance during the first three to four years after
the explosion) until the beginning of active interaction of the
ejecta with the surrounding matter.
Simulations\cite{thielemann90, woosley91} show that the initial
mass of $^{\bf 44}$Ti synthesized in core-collapse supernovae is
{\boldmath $(0.02-2.5)\times 10^{-4}$} solar masses
  (\textit{\textbf{M}}$_{\odot}$).  Hard X-rays and {\boldmath
    $\gamma$}-rays from the decay of this $^{\bf 44}$Ti have
  been unambiguously observed from Cassiopeia A
  only\cite{iyudin94,vink01, renaud06}, leading to the
  suggestion that the values of the initial mass of $^{\bf
    44}$Ti near the upper bound of the predictions occur only in
  exceptional cases\cite{the06}.  For the remnant of supernova
  1987A\cite{arnett89, imshennik89}, an upper limit to the
  initial mass of $^{\bf 44}$Ti of {\boldmath $\leq
    10^{-3}$}\ \textit{\textbf{M}}$_{\odot}$ has been obtained
  from direct X-ray observations\cite{shtykovskiy05}, and an
  estimate of {\boldmath $(1-2)\times 10^{-4}$}
  \ \textit{\textbf{M}}$_{\odot}$ has been made from infrared
  light curves and ultraviolet spectra by complex
  model-dependent computations\cite{chugai97, jerkstrand11,
    larsson11}. Here we report observations of hard X-rays from
  the remnant of supernova 1987A in the narrow band containing
  two direct-escape lines of $^{\bf 44}$Ti at 67.9 and 78.4 keV.
  The measured line fluxes imply that this decay provided
  sufficient energy to power the remnant at late times.  We
  estimate that the initial mass of $^{\bf 44}$Ti was {\boldmath
    $(3.1\pm0.8) \times
    10^{-4}$}\ \textit{\textbf{M}}$_{\odot}$, which is near the
  upper bound of theoretical predictions.\\
\end{abstract}

The only immediate way to determine the initial mass of
$^{44}$Ti (denoted $M_{44}$) in a supernova remnant is through
its emission in the direct-escape lines at energies $E_{i}=
$4.1, 67.9, 78.4, 511 and 1,157 keV.  This isotope decays
according to the chain
$^{44}$Ti$\rightarrow^{44}$Sc$\rightarrow^{44}$Ca, with a
characteristic time of $t_{44}\simeq85.0\pm0.4$ years
(ref. \citen{ahmad06}). Photons in the three X-ray lines, namely
4.1, 67.9 and 78.4 keV, are emitted with efficiencies (average
numbers of photons per decay) $W_i=17.4$\%, 87.7\% and 94.7\%
during the first stage of the decay, and photons in the
$\gamma$-ray lines, namely 511 and 1,157 keV, are produced with
$W_i=188.8$\% and 99.9\% during the second very quick ($\sim
5.7$ h) stage. When the ejecta becomes transparent for
hard-X-ray and $\gamma$-ray lines (typically at $t\ga20$ years),
the corresponding fluxes can be estimated as
$$
F_{i}=\frac{M_{44}W_i}{4\pi\ d^2\ 44\ m_{\rm p} t_{44}}
\exp{(-t/t_{44})} 
$$ where $d$ is the distance to the remnant and $m_{\rm p}$ is
the proton mass.  Measurements\cite{iyudin94, vink01, renaud06}
of the 67.9-, 78.4- and 1,157-keV lines from Cas\,A have yielded
the value $M_{44}=1.6(+0.6/\!-0.3)\times 10^{-4}\ M_{\odot}$:
tentative detections of the 1,157-keV line from the Vela Junior
remnant\cite{iyudin98} and the 4.1-keV line from the G1.9+0.3
remnant\cite{borkowski10} have been also reported, but they are
more uncertain.  Assuming a value of $M_{44}$ similar to that in
Cas~A, we can expect the photon fluxes in these lines from
supernova remnant (SNR) 1987A ($d \simeq 50$ kpc) about $23$
years after the explosion to be about $F_{68}\simeq 0.93
F_{78}\simeq 0.48 F_{68+78}\simeq 0.88 F_{1,157}\simeq 0.46
F_{511}\simeq 3.7\times 10^{-6}$ photons cm$^{-2}$
s$^{-1}$. Note that this remnant was the first for which the
X-ray and $\gamma$-ray emission from the decay of $^{56}$Co
and $^{57}$Co was detected\cite{sunyaev87, matz88,
  sunyaev90}.

The International Gamma-Ray Astrophysics Laboratory
(INTEGRAL)\cite{winkler03} invested considerable observing time
in trying to detect these lines from SNR\,1987A.  The first
observation was carried out in 2003 soon after the launch of
INTEGRAL.  A number of important results have been obtained,
including constraints on the luminosity of the stellar
remnant\cite{shtykovskiy05}, but the dedicated exposure
($\sim1.5$ Ms) was insufficient for the detection of the
$^{44}$Ti lines.  The next set of observations was carried out
in 2010--11 with a total exposure of $\sim 4.5$ Ms.  Figure\,1
presents the results of our analysis of all these data. Images
of the SNR\,1987A field obtained with the IBIS/ISGRI hard-X-ray
telescope on board INTEGRAL are shown in three energy bands,
48--65, 65--82 and 82--99 keV, which encompass the $^{44}$Ti
lines at 67.9 and 78.4 keV. Two known sources, the black-hole
binary \mbox{LMC\,X-1} and the 50-ms Crab-like pulsar
PSR\,B0540-69, are seen in all images. Remarkably, there is an
obvious excess at the position of SNR\,1987A, but only in the
energy band that includes the $^{44}$Ti lines. The location of
the excess coincides within $4^{\prime}$ with that of SNR\,1987A
-- a typical uncertainty for moderately bright sources (the
angular resolution of IBIS is $\simeq12^{\prime}$ full-width at
half-maximum (FWHM)). 

The signal-to-noise ratio for this excess, $S/N\simeq 4.1$
(related to the flux measured in both lines at once),
corresponds to a rather small probability of detecting such an
excess by chance, $P_0\simeq2\times10^{-5}$. The probability of
detecting a random excess with the same $S/N$ ratio at an
arbitrary position in the whole $29^{\circ}\times29^{\circ}$
IBIS field of view is much higher (it can be estimated as $m
P_0\simeq 0.43$ where $m\simeq (29^{\circ}/12^{\prime})^2
\simeq2.1\times10^4$ is the number of uncorrelated spots within
the IBIS field of view; a `spot' is the area of the sky image
that can be occupied by a single source). In reality, the
probability of detecting the $^{44}$Ti lines by chance is even
smaller than $P_0$, because significant excesses were
simultaneously detected at the position of SNR\,1987A in the
images accumulated in the narrow energy bands 62.7--73.2 and
73.2--83.7 keV. Each of these bands contains one of the
$^{44}$Ti lines and they are nearly independent in terms of the
energy resolution of IBIS/ISGRI, which is 8\% FWHM at $\sim70$
keV. The $S/N$ ratios measured for SNR\,1987A in these images
are respectively 3.0 and 3.1, and the corresponding chance
probabilities are $P_1=1.3\times10^{-3}$ and
$P_2=9.7\times10^{-4}$. The combined probability of detecting
the $^{44}$Ti emission by chance is then equal to $P_1\times
P_2\simeq1.3\times10^{-6}$ ($\simeq4.7\sigma$).

The hard-X-ray spectrum obtained with IBIS/ISGRI from SNR\,1987A
is presented in Fig.\,2.  We fitted it using two lines of
Gaussian shape with fixed centroid energies, equal widths and
self-consistent (according to the $^{44}$Ti emission
efficiencies) normalizations. The flux measured in both lines at
once, $F_{68+78}\simeq(1.7\pm0.4)\times10^{-5}\ \mbox{photons
  cm}^{-2}\ \mbox{s}^{-1}$, corresponds to a value for initial
amount of synthesized $^{44}$Ti of $M_{44}\simeq
(3.5\pm0.8)\times 10^{-4}\ M_{\odot}$ (the uncertainty is purely
statistical).  This estimate can be slightly decreased if we
assume the presence of some underlying continuum in the
SNR\,1987A spectrum -- for example, in the form of a power law
with a photon index of 2.1 (a Crab-like spectrum), which should
describe the hard-X-ray emission well up to $\sim80$ keV
(ref. \citen{berezhko11}). Its possible contribution at 1 keV,
$(8.5\pm3.2)\times10^{-4}\ \mbox{photons cm}^{-2}\ \mbox{s}^{-1}
\mbox{keV}^{-1}$, coincides with the normalization of the
non-thermal X-ray component in the SNR\,1987A spectrum following
from the XMM-Newton observations in 2009\cite{sturm10}. With
such a component (shown in Fig.\,2 by a solid curve) the flux in
the lines decreases to
$F_{68+78}=(1.5\pm0.4)\times10^{-5}\ \mbox{photons
  cm}^{-2}\ \mbox{s}^{-1}$, which may be translated to
$M_{44}\simeq (3.1\pm0.8) \times 10^{-4}\ M_{\odot}$.

The SPI $\gamma$-ray spectrometer on board INTEGRAL observed
SNR\,1987A simultaneously with IBIS/ISGRI in the 511- and
1,157-keV lines of $^{44}$Ti; however, the smaller effective area
and 
presence of strong background features at 511 and $\sim 1,100$ keV
preclude their confident detection. The photon spectrum measured
in the 1,106--1,205-keV band is presented in Fig.\,3.  
The solid curve shows the $1.7\sigma$ upper limit (90\%
confidence) for the flux in the 1,157-keV Gaussian line, namely $F_{1,157}\leq2.6 \times10^{-5}\ \mbox{photons  cm}^{-2}$ s$^{-1}$.
It can be converted to the limit for the amount of $^{44}$Ti,
$M_{44}\la9 \times10^{-4}\ M_{\odot}$.
For comparison, the dotted curve in Fig.\,3 shows the Gaussian
line corresponding to the value of $M_{44}$ measured with IBIS/ISGRI.

The detection of emission in the 67.9- and 78.4-keV lines allows
us to measure directly the amount of radioactive $^{44}$Ti
synthesized during the explosion, that is, $M_{44}\simeq
(3.1\pm0.8) \times 10^{-4}\ M_{\odot}$. This amount formally
exceeds by a factor of 1.5 to 2 the estimates based on
theoretical simulations of explosive nucleosynthesis in
supernova 1987A and on the infrared and ultraviolet data
(although the discrepancy represents only one standard
deviation). We note that $^{44}$Ti may be produced either as a
result of strong deviation from nuclear statistical equilibrium
during the explosion (so-called $\alpha$-rich
freeze-out)\cite{woosley91} or as a result of incomplete
Si-burning in a thin shell surrounding the region of main energy
release\cite{woosley91, thielemann96}. In both cases, the
simulations are not straightforward and many poorly known
parameters (for example, the mass of the collapsed core, the
maximum temperature and density behind the shock wave, the
neutron excess and the composition of the pre-supernova
interior) could affect the results. In this sense, the value of
$M_{44}$ that we report here may be useful for studying
explosive nucleosynthesis and physical conditions reached during
the shock-wave passage in the ejecta.

We note finally that the shock interaction of the ejecta and the
surrounding matter is responsible for the synchrotron emission
observed at present from SNR\,1987A in the radio- and soft-X-ray
bands\cite{park05}.  The reprocessing of this emission in the
ejecta is contributing to the late-time optical-infrared
emission of SNR\,1987A\cite{larsson11}.  Our estimate of the
power-law component in the spectrum of SNR\,1987A may be
translated to a $2\sigma$ upper limit for the luminosity of
synchrotron emission of $\simeq3\times10^{35}\ \mbox{erg
  s}^{-1}$ in the 20--60 keV band. This value can be also
considered as the limit for the luminosity of the stellar
remnant (pulsar or black hole) harboured inside the ejecta.

\parskip 8pt

\noindent
{\sl Received 23 November 2011; accepted 13 August 2012;
  published 18 October 2012.}\\

\begin{addendum}\baselineskip 20pt
 \item We thank R.A. Sunyaev for reading of the manuscript and for
   comments; V.S. Imshennik, D.K. Nadezhin and N.N. Chugai for
   discussions about different aspects of the physics of the
   supernova\,1987A explosion; R.A. Krivonos for information regarding
   the analysis of INTEGRAL/IBIS data; and M. Coe for allowing
   us access to his INTEGRAL data of the LMC observations ($\sim 1$
   Ms) before the end of the proprietary period.  This work is
   based on data obtained through the Russian and European
   INTEGRAL science data centers, and was supported by grants
   RFBR-11-02-12285ofi-m-2011 and RAS-P20 `The origin,
   structure and evolution of objects of the Universe'.

\item[Author Contributions] S.A.G. was Principal Investigator of the
  proposal requesting INTEGRAL time ($\sim 3.5$ Ms) to observe
  SNR\,1987A; made a preliminary analysis of the IBIS/SIGRI data
  and detected an excess at the SNR\,1987A position; and wrote a
  draft of the text. A.A.L. was Co-Investigator of the proposal,
  and participated in the analysis of IBIS/ISGRI data and the
  verification of the result. S.S.T.  carried out the analysis
  of SPI data. C.W. provided general support for this project
  and participated in its different aspects, and C.W. and
  A.A.L. contributed substantially to the final text. All
  authors discussed the results and their presentation.

 \item[Author Information] Reprints and permissions information
   is available at www.nature.com/reprints. The authors declare
   no competing financial interests. Reader are welcome to
   comment on the online version of the paper.  Correspondence
   and requests for materials should be addressed to
   S.A.G. (sergei@hea.iki.rssi.ru).
\end{addendum}
\clearpage
\baselineskip 16pt 
\begin{figure}
\centering
\begin{minipage}[t]{9cm}
\epsfig{file=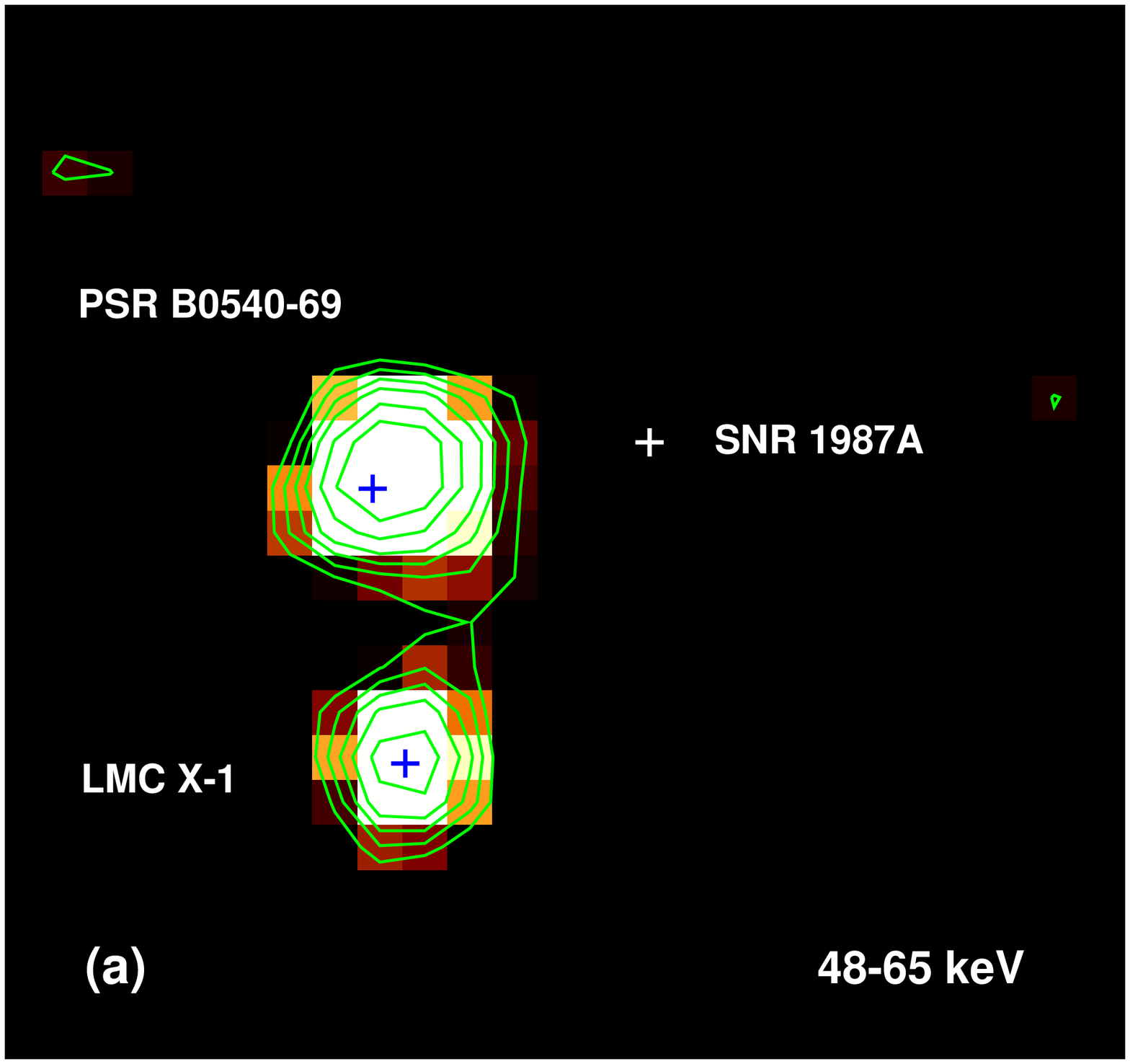,width=0.97\linewidth}\\
\epsfig{file=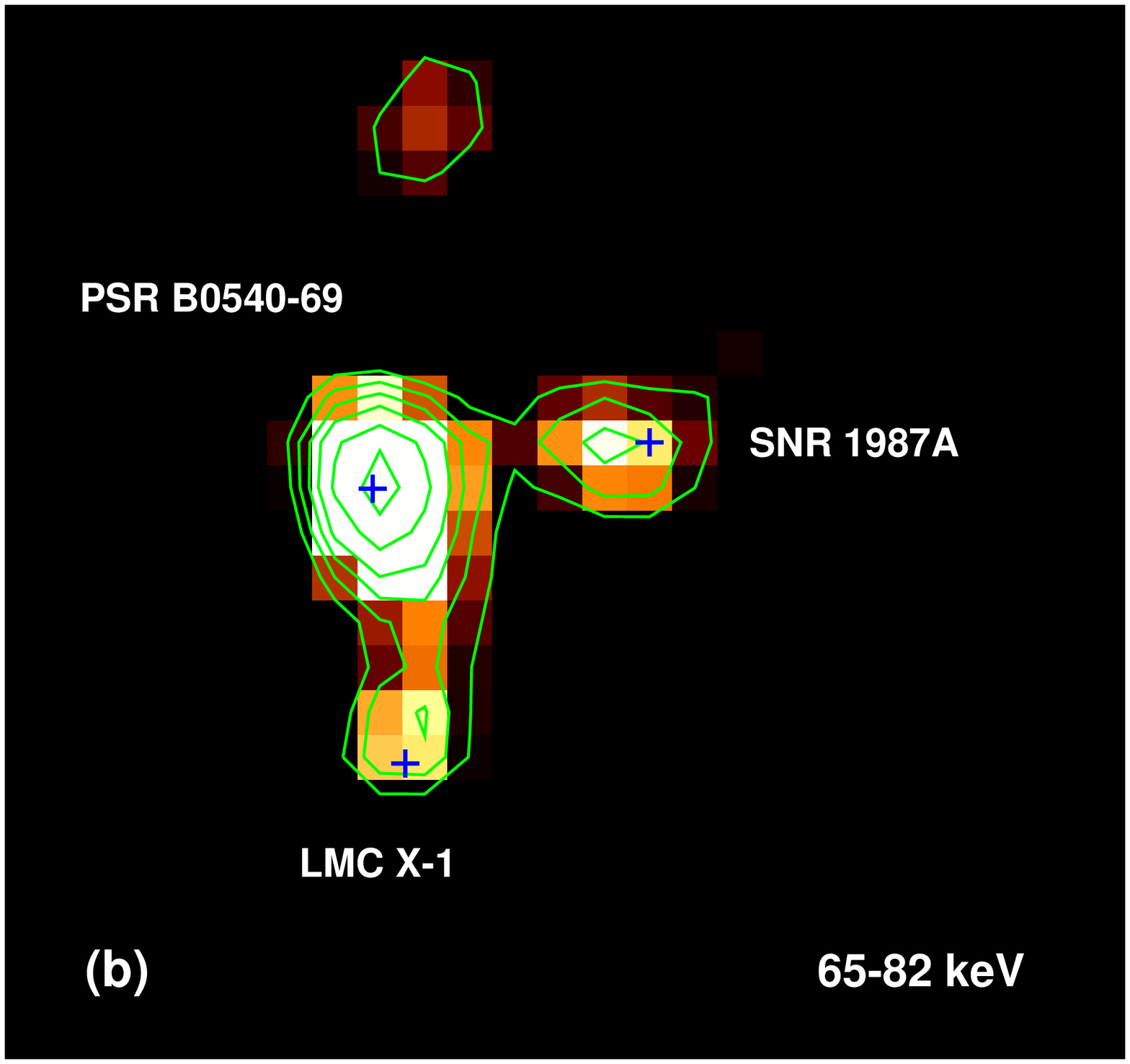,width=0.97\linewidth}\\
\epsfig{file=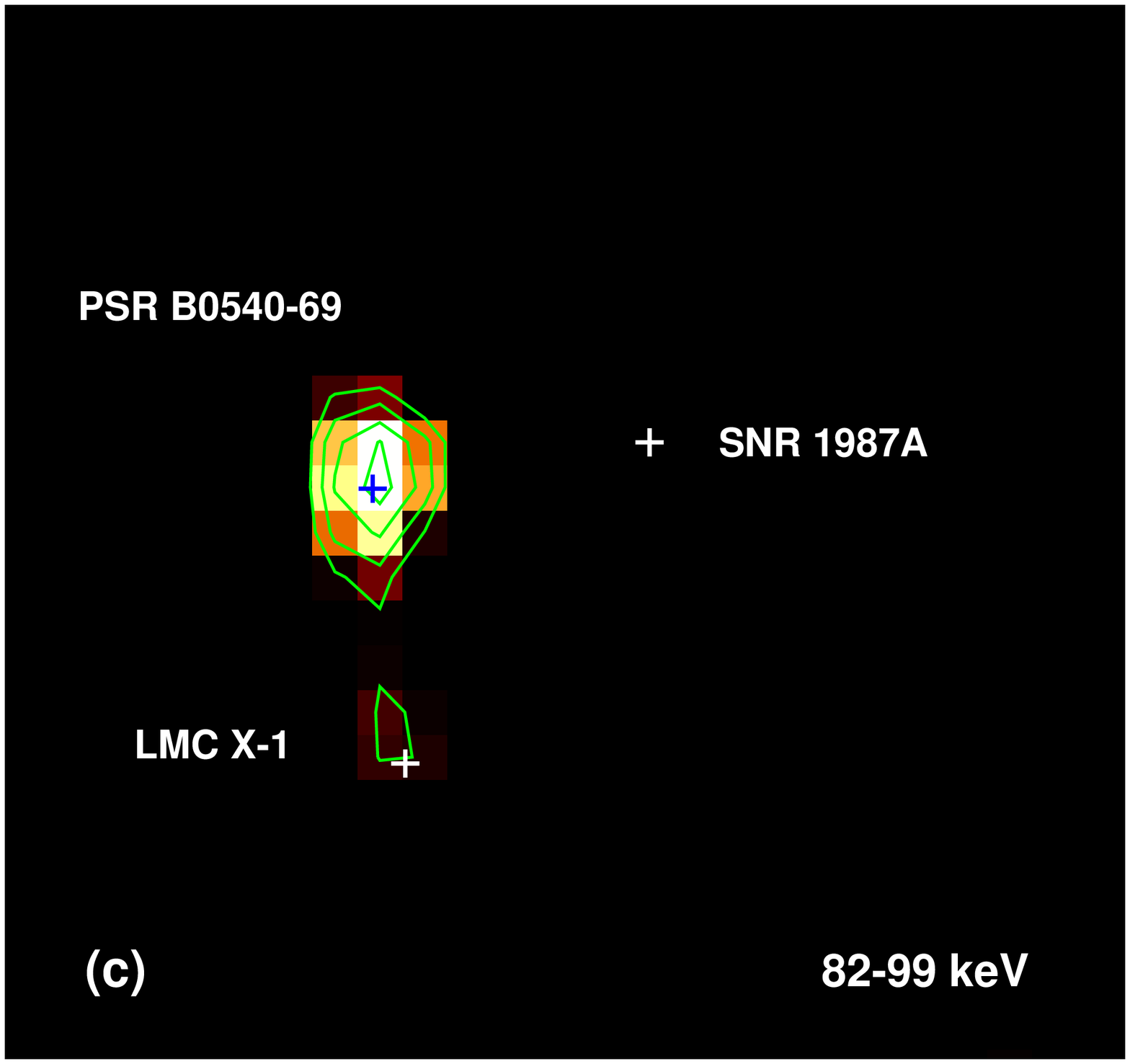,width=0.97\linewidth}
\end{minipage} \hspace{1em} \begin{minipage}[t]{7cm}
\caption{\baselineskip 16pt \textbf{Hard X-ray images indicating the detection of
    $^{44}$Ti emission lines from SNR\,1987A. a-c,} Maps of the
  signal-to-noise ratio ($S/N$) of the $1.^{\!\!\!\circ}5
  \times1.^{\!\!\!\circ}5$ sky region around SNR\,1987A
  accumulated in three energy bands with the IBIS/ISGRI
  telescope on board INTEGRAL during observations in 2003--2011
  ($\sim 6.0$ Ms of real exposure or $\sim 4.2$ Ms of
  dead-time-corrected exposure): 48--65 keV (\textbf{a}); 65--82
  keV (\textbf{b}); 82--99 keV (\textbf{c}); The maps were
  reconstructed using standard technique\cite{krivonos10} with
  contours given at $S/N$ levels of 2.7, 3.3, 3.9, 4.5, 5.4 and
  6.3. Two well-known sources, PSR B0540-69 and LMC\,X-1, are
  seen bright in all three images, but SNR\,1987A is confidently
  detected only in \textbf{b,} in the band that contains the 67.9-
  and 78.4-keV direct-escape lines of radioactive $^{44}$Ti
  decaying inside the ejecta.}
\end{minipage}
\label{images}
\end{figure}
\clearpage
\begin{figure}
\centering
\epsfig{file=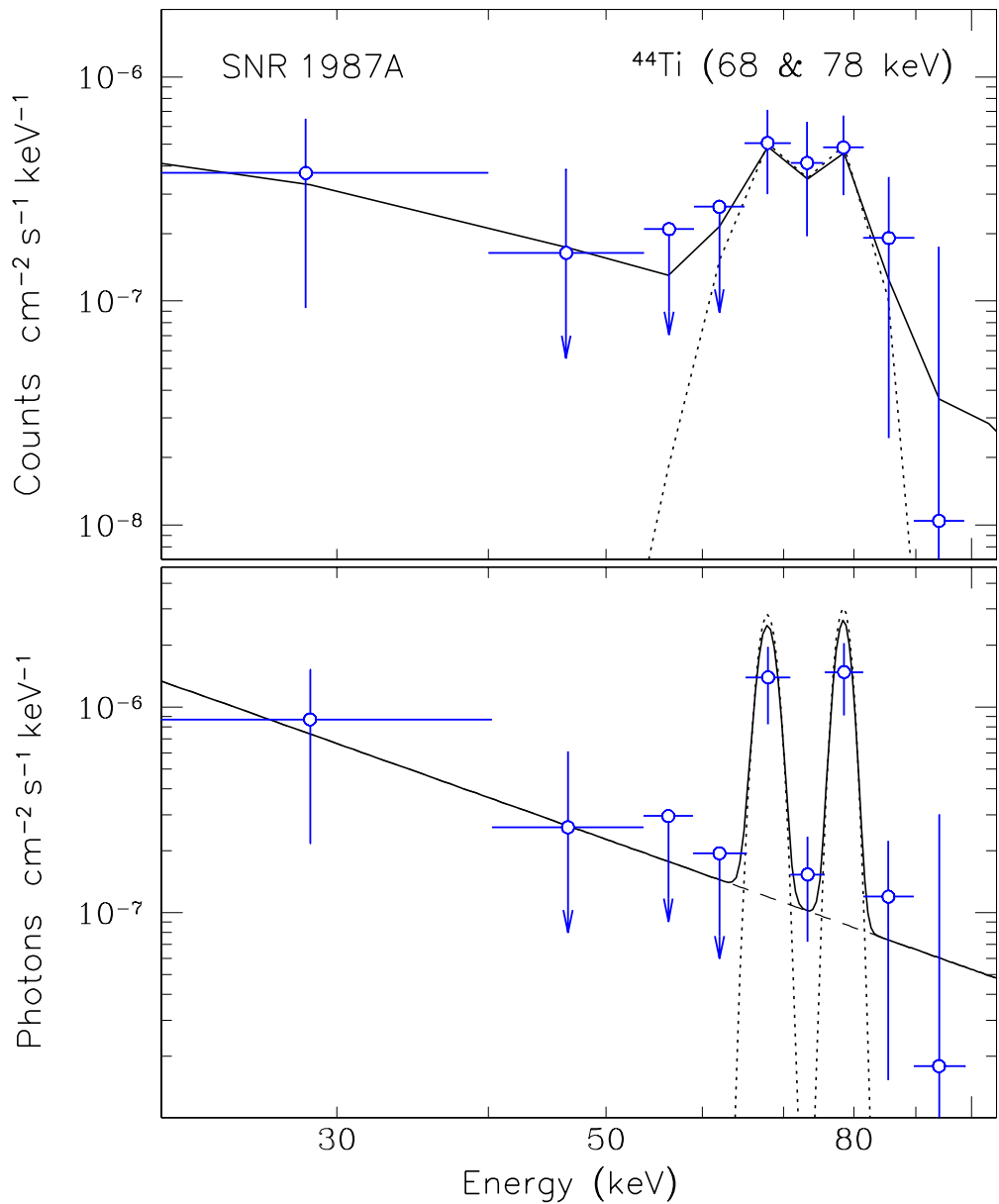,width=0.86\linewidth}

\caption{\textbf{Hard X-ray spectrum of SNR\,1987A measured with
    IBIS/ISGRI in 2003--2011.  a,} The observed (count) spectrum
  together with the simulated response curves; \textbf{b,} the
  unfolded (photon) spectrum. In each panel, open circles show
  the spectra, dotted curves show the fit with two Gaussian
  lines at 67.9 and 78.4 keV of the radioactive decay of
  $^{44}$Ti and solid curves show a similar fit which includes
  an additional power-law continuum with photon index
  $\alpha=2.1$. Each data point (open circles) represents the
  result of flux measurement at the position of SNR\,1987A in
  the image corresponding to the given energy band. It is
  expected that $^{44}$Ti is produced inside the supernova core,
  which is expanding with velocity $v_{44}\leq1,700$ km
  s$^{-1}$; thus, the internal width of the lines (that is,
  their width before smoothing due to the finite energy
  resolution of the instrument) is unlikely to exceed $\Delta
  E\simeq 0.4$ keV. We kept the width of the lines used for
  fitting fixed at $2.8$ keV FWHM, which is half of the energy
  resolution of ISGRI (such choice does not affect flux
  measurements in the internally narrow lines, thereby making it
  possible to work with the discrete response matrix of
  ISGRI). The upper limits are $1.7 \sigma$ (90\% confidence),
  error bars on the other points are $1\sigma$.}
\label{spectrum1}
\end{figure}
\clearpage
\begin{figure}
\centering
\epsfig{file=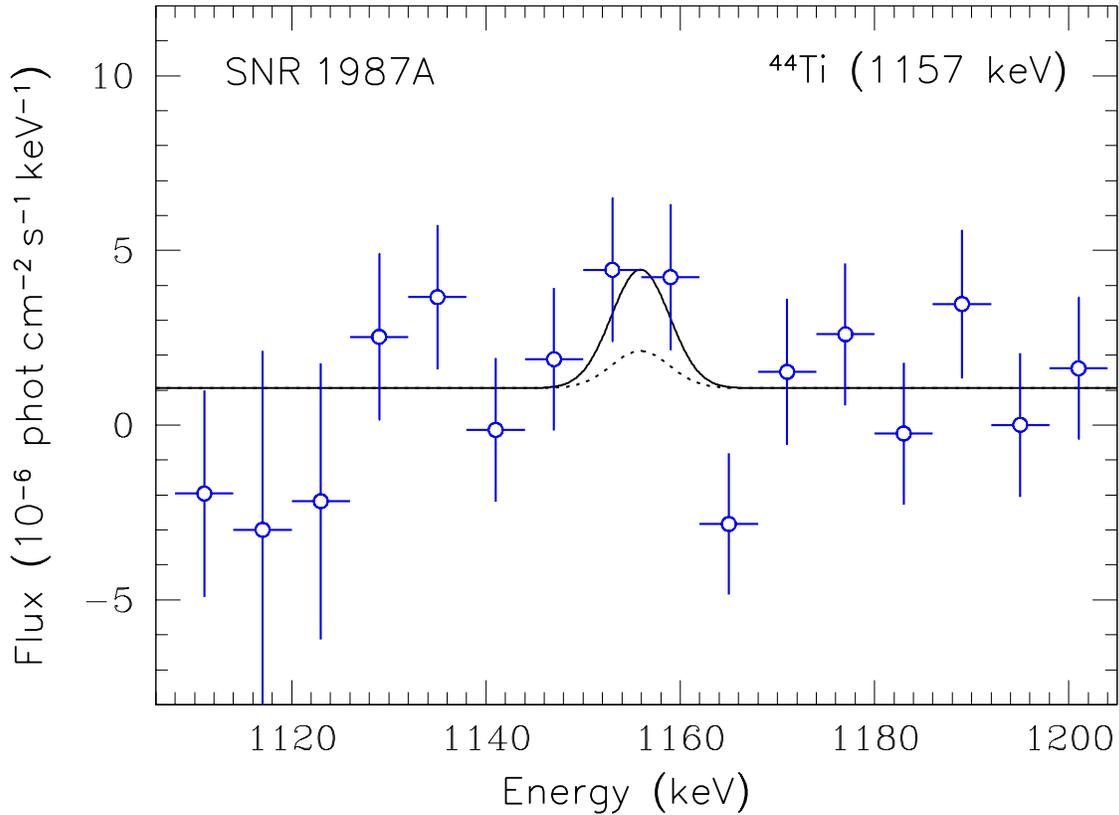,width=0.96\linewidth}

\caption{{\boldmath $\gamma$}\textbf{-ray spectrum of SNR\,1987A measured with
    SPI in 2003--2011.}  The spectrum was reconstructed using
  standard techniques\cite{chur11} near the highest-energy
  (1,157 keV) direct-escape line of $^{44}$Ti.  The solid curve
  shows the Gaussian line with the flux at its $1.7\sigma$ upper
  limit. Its centroid energy was redshifted by 1.1 keV to take
  into account the heliocentric velocity of SNR\,1987A (285 km
  s$^{-1}$; ref. \citen{meaburn95}); its FWHM was taken to be
  7.0 keV (internal width $\sim 1,157\ \mbox{keV} \times
  v_{44}/c\simeq6.5\ \mbox{\rm keV},$ corrected for the energy
  resolution of SPI, namely $\sim 3$ keV FWHM at 1.2 MeV; here
  $v_{44}\leq1700\ \mbox{km s}^{-1}$ is the expansion velocity
  of the ejecta layer containing $^{44}$Ti, and $c$ is the
  velocity of light).  The similar Gaussian line with total flux
  corresponding to the $^{44}$Ti mass measured by IBIS/ISGRI in
  the hard-X-ray lines is shown by a dotted curve. Error bars
  are $1\sigma$.}
\label{spectrum2}
\end{figure}

\end{document}